\documentclass[twocolumn,prc,showpacs,showkeys]{revtex4}
\usepackage{amsfonts}
\usepackage{amsmath}
\usepackage{amssymb}
\usepackage{graphicx}
\usepackage{rotating}

\providecommand{\U}[1]{\protect\rule{.1in}{.1in}}
\providecommand{\U}[1]{\protect\rule{.1in}{.1in}}
\providecommand{\U}[1]{\protect\rule{.1in}{.1in}}

\begin{document}

\preprint{}
\title{Electron assisted $dd$ reactions in metals\\
}
\author{P\'{e}ter K\'{a}lm\'{a}n\footnote{%
retired from Budapest University of Technology and Economics, Institute of
Physics, \newline
e-mail: kalmanpeter3@gmail.com}}
\author{Tam\'{a}s Keszthelyi}
\affiliation{Budapest University of Technology and Economics, Institute of Physics,
Budafoki \'{u}t 8. F., H-1521 Budapest, Hungary\ }

\begin{abstract}
The electron assisted low energy $dd$ reactions in deuterized metals are
investigated. It is shown that if a metal is irradiated with slow, free
deuterons then the $e+d+d\rightarrow e^{\prime }+p+t$ and $e+d+d\rightarrow
e^{\prime }+n+$ $^{3}He$\ electron assisted $dd$ processes have measurable
probabilities even in the case of slow deuterons. The cross sections and the
yields in an irradiated sample are determined. The results are connected
with the so called anomalous screening effect. It is concluded that the
electron $dd$ processes have to be taken into account when evaluating the
experimental data of low energy fusion reactions in metals.
\end{abstract}

\startpage{1}
\endpage{}
\maketitle

Astrophysical factors have great importance in nuclear astrophysics \cite%
{Angulo}, \cite{Descouvemont}. In the last two decades, investigating
astrophysical factors of nuclear reactions of low atomic numbers, in the
cross section measurements of the $dd$ reactions in deuterated metal targets
extraordinary observations were made in low energy accelerator physics \cite%
{Raiola1}. The phenomenon of increasing cross sections of the reactions
measured in solids compared to the cross sections obtained in gaseous
targets is the so called anomalous screening effect. A few years ago a
systematical survey of the experimental methods applied in investigating and
the theoretical efforts for the explanation of the anomalous screening
effect was done \cite{Huke} from which one can conclude that the full
theoretical explanation of the effect is still open. Here it is shown that
the electron assisted $dd$ processes in metal environment produce yields
which are considerable in the energy range investigated in \cite{Huke}.

Let us consider the following $dd$ nuclear reactions

\begin{equation}
d+d\rightarrow \text{ }p+t+\Delta ,  \label{R2}
\end{equation}

\begin{equation}
d+d\rightarrow n+\text{ }^{3}He+\Delta .  \label{R3}
\end{equation}%
(Since the work \cite{Huke} focuses to reactions $\left( \ref{R2}\right) $
and $\left( \ref{R3}\right) $ the $d+d\rightarrow $ $^{4}He+\gamma $
reaction is not discussed in this paper.) Here $p$, $d$ and $t$ denote
proton, deuteron and triton, respectively, and $\Delta $ is the energy of
the reaction, i.e. the difference between the rest energies of the initial $%
\left( d+d\right) $ and final $\left( \text{ }p+t\text{ and }n+\text{ }%
^{3}He\right) $ states, $\Delta (pt)=$ $4.033$ $MeV$ and $\Delta (nHe)=3.269$
$MeV$, respectively.

The energy dependence of the cross section $\left( \sigma \right) $ of the
processes $\left( \ref{R2}\right) $ and $\left( \ref{R3}\right) $ reads as 
\begin{equation}
\sigma \left( E_{r}\right) =S\left( E_{r}\right) \exp \left[ -2\pi \eta
\left( E_{r}\right) \right] /E_{r},  \label{SigmaE}
\end{equation}%
where $E_{r}$ is the relative kinetic energy of the deuterons, $S\left(
E_{r}\right) $ is the astrophysical factor and $\eta \left( E_{r}\right) $
is the Sommerfeld parameter, 
\begin{equation}
\eta \left( E_{r}\right) =\alpha _{f}\frac{m_{0}c}{\hbar k\left(
E_{r}\right) }=\alpha _{f}\sqrt{\frac{m_{0}c^{2}}{2E_{r}}},  \label{etak}
\end{equation}%
where $k$ is the magnitude of the wave vector of the relative motion of the
interacting deuterons in the center of mass coordinate system. The reduced
rest mass $m_{d}/2$ of two deuterons of rest mass $m_{d}$ is approximated as 
$m_{d}/2=m_{0}c^{2}=931.494$ $MeV$ which is the atomic mass (energy) unit. $%
\hbar$ is the reduced Planck constant, $c$ is the velocity of light and $%
\alpha _{f}$ is the fine structure constant \cite{Angulo}, \cite%
{Descouvemont}. In the laboratory frame of reference the energy $E$\ of the
deuteron in the beam is $E=2E_{r}$.

The $\exp \left[ -2\pi \eta \left( E_{r}\right) \right] $ dependence of $%
\sigma \left( E_{r}\right) $ appears since the relative motion of the two
deuterons is described by the Coulomb solution $\varphi (\mathbf{r})$ $\sim
e^{-\pi \eta /2}\Gamma (1+i\eta )$, and the cross section of the process is
proportional to 
\begin{equation}
\left\vert e^{-\pi \eta /2}\Gamma (1+i\eta )\right\vert ^{2}=\frac{2\pi \eta
\left( E_{r}\right) }{\exp \left[ 2\pi \eta \left( E_{r}\right) \right] -1}%
=F_{C}(E_{r}).  \label{Fjk}
\end{equation}%
Here $\Gamma $ is the Gamma function \cite{Alder}. $F_{C}\left( E_{r}\right) 
$ is the square of the deuteron-deuteron Coulomb factor. The rate of nuclear
reactions $\left( \ref{R2}\right) $ and $\left( \ref{R3}\right) $ decreases
rapidly with decreasing beam energies in consequence of $F_{C}\left(
E_{r}\right) $ becoming small.

However the situation changes in the case of the electron assisted versions
of $\left( \ref{R2}\right) $ and $\left( \ref{R3}\right) $ which read%
\begin{equation}
e+d+d\rightarrow e^{\prime }+p+t+\Delta ,  \label{R2e}
\end{equation}%
\begin{equation}
e+d+d\rightarrow e^{\prime }+n+\text{ }^{3}He+\Delta ,  \label{R3e}
\end{equation}%
where $e$ denotes a conduction electron of the metal and $e^{\prime }$
stands for it after assisting the reaction. The cross section of reactions $%
\left( \ref{R2e}\right) $ and $\left( \ref{R3e}\right) $\ is proportional to
the square of the electron-deuteron Coulomb factor expressed in the variable 
$E$ 
\begin{equation}
F_{e}\left( E\right) =\left\vert \frac{2\pi \eta _{e}\left( E\right) }{\exp %
\left[ 2\pi \eta _{e}\left( E\right) \right] -1}\right\vert  \label{Fek}
\end{equation}%
where 
\begin{equation}
\eta _{e}\left( E\right) =-\alpha _{f}\frac{m_{e}c}{\hbar k_{r}\left(
E\right) }=-\alpha _{f}\sqrt{\frac{m_{0}c^{2}}{E}}  \label{etaek}
\end{equation}%
with $m_{e}$ the electronic rest mass. $k_{r}$ is the magnitude of the wave
vector of motion in the center of mass coordinate system of the interacting
deuteron and electron. Here the reduced mass $\mu =m_{e}m_{d}/\left(
m_{e}+m_{d}\right) $ is approximated as $\mu =m_{e}$. In the cases $(5$ $%
keV<E<50$ $keV)$ investigated \cite{Huke} $F_{e}\left( E\right) )=\left\vert
2\pi \eta _{e}\left( E\right) \right\vert $. Although the processes $\left( %
\ref{R2e}\right) $ and $\left( \ref{R3e}\right) $ are second order processes
it is expected that the disappearance of the small $\exp \left[ -2\pi \eta
\left( E\right) \right] $ quantity from the cross section of processes $%
\left( \ref{R2e}\right) $ and $\left( \ref{R3e}\right) $ may make the order
of magnitude of their cross sections become comparable with or higher than
the order of magnitude of $\left( \ref{SigmaE}\right) $ with decreasing $E$.

In the processes investigated a low energy deuteron beam bombs deuterized
metal targets. The interaction between an ingoing deuteron and an electron
of the metal is a Coulomb interaction of potential $V^{Cb}\left( \mathbf{x}%
\right) =\int -\left( 2\pi ^{2}\right) ^{-1}e^{2}\left(
q^{2}+q_{s}^{2}\right) ^{-1}\exp \left( i\mathbf{q}\cdot \mathbf{x}\right) d%
\mathbf{q}$ with coupling strength $e^{2}=\alpha _{f}\hbar c$ and $q_{s}=%
\sqrt{4\pi e^{2}\varrho \left( \varepsilon _{F}\right) }$ the Thomas-Fermi
screening parameter, where $\varrho \left( \varepsilon _{F}\right) $ is the
density of electron states at the Fermi-energy $\varepsilon _{F}$ in the
metal \cite{Solyom}. The order of magnitude of $q_{s}$ is $10^{8}$ $cm^{-1}$%
~for metallic elements. The other interaction that is taken into account
between the nucleons of the deuterons is the strong interaction of potential 
$V^{St}\left( \mathbf{x}\right) =-f\exp \left( -s\left\vert \mathbf{x}%
\right\vert \right) /\left\vert \mathbf{x}\right\vert $, where the strong
coupling strength $f=0.08\hbar c$ \cite{Bjorken} and $1/s$ is the range of
the strong interaction.

In the second order process investigated a free electron of the conduction
band of the metal takes part in a Coulomb scattering with an ingoing
deuteron of the beam. In the intermediate state a virtual free
neutron-proton pair is created in the $\left( \ref{R2e}\right) $ and $\left( %
\ref{R3e}\right) $ processes. The virtual neutron or proton corresponding to
the processes $\left( \ref{R2e}\right) $ and $\left( \ref{R3e}\right) $,
respectively, is captured due to the strong interaction by an other deuteron
located in the solid. The reaction energy $\Delta $ is shared between the
quasi-free final electron and the proton and tritium or neutron and $^{3}He$
corresponding to the processes $\left( \ref{R2e}\right) $ and $\left( \ref%
{R3e}\right) $, respectively. Since the aim of this paper is to show the
fundamentals of the main effects, the simplest description is chosen.

The transition probability per unit time $\left( W_{fi}\right) $ of the
processes can be written as $W_{fi}=\frac{2\pi }{\hbar }\sum_{f}\left\vert
T_{fi}\right\vert ^{2}\delta (E_{f}-\Delta )$ with $T_{fi}=\sum_{\mu
}V_{f\mu }^{St}V_{\mu i}^{Cb}/\Delta E_{\mu i}$. Here $V_{\mu i}^{Cb}$ \ is
the matrix element of the Coulomb potential between the initial and
intermediate states and $V_{f\mu }^{St}$ is the matrix element of the
potential of the strong interaction between intermediate and final states,
furthermore $\Delta E_{\mu i}=E_{\mu }-E_{i}+\Delta _{\mu i}$. $E_{i}$, $%
E_{\mu }$ and $E_{f}$ are the kinetic energies in the initial, intermediate
and final states, respectively. $\Delta _{\mu i}$ is the difference between
the rest energies of the intermediate and initial states.

In reactions $\left( \ref{R2e}\right) $, $\left( \ref{R3e}\right) $ the
Coulomb interaction virtually breaks up the deuteron and produces a
virtually free proton and neutron. Correspondingly,~$\Delta _{\mu i}=2.225$ $%
MeV$ which is the binding energy of the deuteron. The method of evaluation
of matrix elements $V_{\mu i}^{Cb}$, $V_{f\mu }^{St}$ and the transition
probability per unit time is similar to the method discussed recently \cite%
{kk3} determining the cross section of electron assisted neutron exchange
process.

The cross section $\sigma _{pt}$ of the electron assisted $d+d\rightarrow p+t
$ process has considerable contributions if%
\begin{equation}
\ \left\vert \mathbf{k}_{1}+\mathbf{k}_{2}\right\vert \lesssim q_{s}
\label{condlambda}
\end{equation}%
where $\mathbf{k}_{1}$\textbf{,\ }$\mathbf{k}_{2}$ are the wave vectors in
the final state of the proton and the triton. In the cases which fulfill
condition $\left( \ref{condlambda}\right) $ $\mathbf{k}_{1}=-\mathbf{k}%
_{2}+\delta \mathbf{k}$ with $\left\vert \delta \mathbf{k}\right\vert
=\delta k\sim q_{s}$ and $k_{1}\simeq k_{2}\simeq k_{0}=\sqrt{\left(
3/2\right) m_{0}\Delta }/\hbar $, which is determined by the energy Dirac
delta, resulting $\delta k/k_{0}\sim q_{s}/k_{0}\ll 1$. Evaluating $\sigma
_{pt}$ the Weisskopf and the long wavelength approximations are used, i.e.
for the initial\ and final bound neutron states we take $\Phi _{W}\left( 
\mathbf{r}\right) =\sqrt{3/\left( 4\pi R^{3}\right) }$, if $\left\vert 
\mathbf{r}\right\vert \leq R$ and $\Phi _{W}\left( \mathbf{r}\right) =0$ for 
$\left\vert \mathbf{r}\right\vert >R$, $R=\{R_{d}$ or $R_{t}\}$ where $%
R_{d}=4.31$ $fm$ is the radius of the deuteron and $R_{t}\simeq
R_{^{3}He}=2.26$ $fm$ is the radius of the triton \cite{Blatt} and $\exp
\left( -i\mathbf{k}_{0}\cdot \mathbf{r}\right) =1$ is used if $\left\vert 
\mathbf{r}\right\vert \leq R$. Furthermore $s=1/r_{0}$ with $r_{0}=1.2\times
10^{-13}$ $cm$ is applied. The further details of the calculation are
similar to those made in \cite{kk3}. So%
\begin{equation}
\sigma _{pt}=C_{pt}\frac{u}{E}  \label{sigma2}
\end{equation}%
with $E$ the kinetic energy of the deuterons in the beam, $u$ the deuteron
over metal number densities and%
\begin{equation}
C_{pt}=\frac{K_{0}}{\left[ 1+\frac{2\Delta _{\mu i}}{3\Delta \left(
pt\right) }\right] ^{2}}\frac{N_{c}}{v_{c}q_{s}\Delta \left( pt\right) },
\label{Cpt}
\end{equation}%
where $v_{c}$ is the volume of the elementary cell of the metal, $N_{c}$ is
the number of lattice sites in the elementary cell and $K_{0}=0.598\frac{%
2^{12}}{3^{6}}\pi ^{2}\alpha _{f}^{2}\left( 0.08\right) ^{2}\left(
R_{d}/R_{t}\right) ^{3}r_{0}^{4}\left( m_{0}c^{2}\right) ^{2}$.

The cross section $\sigma _{nHe}$\ of the electron assisted $d+d\rightarrow
n+$ $^{3}He$ reaction reads%
\begin{equation}
\sigma _{nHe}=C_{nHe}\frac{u}{E}  \label{sigma3}
\end{equation}%
where%
\begin{equation}
C_{nHe}=\frac{K_{0}}{\left[ 1+\frac{2\Delta _{\mu i}}{3\Delta \left(
nHe\right) }\right] ^{2}}\frac{N_{c}F_{pd}}{v_{c}q_{s}\Delta \left(
nHe\right) }  \label{CnHe}
\end{equation}%
with $F_{pd}=2\pi \eta _{pd}/\left( e^{2\pi \eta _{pd}}-1\right) $, which is
the square of the $pd$ Coulomb factor before the action of strong
interaction with $\eta _{pd}=\alpha _{f}\sqrt{2m_{0}c^{2}/\left( 3\Delta
\left( nHe\right) \right) }$.

The yield $\left( dN/dt\right) _{pt}$ of events of electron assisted $%
d+d\rightarrow p+t$ process in an elementary small volume $V_{s}$ of the
sample can be written as $\left( dN/dt\right) _{pt}=N_{t}\sigma _{pt}\Phi $,
where $\Phi $ is the flux of deuterons, $N_{t}$ is the number of target
particles, i.e. the number of free electrons of the conduction band of the
metal. The number of elementary cells in an elementary small volume $V_{s}$
of the sample is $V_{s}/v_{c}$ and the number of conduction electrons in an
elementary cell is $g_{e}$ thus the number $N_{t}=g_{e}V_{s}/v_{c}$. Using $%
\left( \ref{sigma2}\right) $\ the yield reads as $\left( dN/dt\right)
_{pt}=g_{e}V_{s}C_{pt}u\Phi /\left( v_{c}E\right) $, where the quantities $%
\Delta \left( pt\right) $ and $E$ have to be substituted in $MeV$ units. The
yield of the events of the electron assisted $d+d\rightarrow n+$ $^{3}He$
reaction produced in an elementary small volume $V_{s}$ of the sample reads
as $\left( dN/dt\right) _{nHe}=g_{e}V_{s}C_{nHe}u\Phi /\left( v_{c}E\right) $%
.

The yield $\left( dN/dt\right) _{usual}$ of the usual $d+d\rightarrow p+t$
process (without taking into account screening) can be written as $\left(
dN/dt\right) _{usual}=N_{c}V_{s}\sigma \left( E\right) u\Phi /v_{c}$ where $%
N_{c}$ is the number of atoms in the elementary cell and $\sigma \left(
E\right) =\left( 2S\left( 0\right) /E\right) \exp \left[ -2\pi \eta \left(
E\right) \right] $ with $\eta \left( E\right) =\alpha _{f}\sqrt{m_{0}c^{2}/E}
$. Here $\sigma \left( E\right) $ is expressed in the variable $E=2E_{r}$.
In the extremely low energy range the $S(E_{r})=S(0)$ approximation is valid
and the $S(0)$ values of processes $\left( \ref{R2}\right) $ and $\left( \ref%
{R3}\right) $ are: $5.6\times 10^{-2}$ and $5.5\times 10^{-2}$ in $MeVb$
units \cite{Angulo}.

It is useful to introduce the relative yield 
\begin{equation}
r=\frac{\left( \frac{dN}{dt}\right) _{pt}}{\left( \frac{dN}{dt}\right)
_{usual}}=\frac{g_{e}C_{pt}}{2N_{c}S(0)}\exp \left[ 2\pi \eta \left(
E\right) \right] ,  \label{ratio}
\end{equation}%
which is the ratio of the yields of electron assisted and normal $%
d+d\rightarrow p+t$ processes in an elementary volume of the sample. As
model material we take $Pd$. It has $v_{c}=d^{3}/4$ with $d=3,89\times
10^{-8}$ $cm$ and $N_{c}=2$.\ Using the data $\rho \left( \varepsilon
_{F}\right) =25$ $states/atom/Rydberg$ for the density of states at the
Fermi energy of $Pd$ at $u=0.3$ \cite{Huke}, \cite{Faulkner}, one obtains $%
q_{s}=6.73\times 10^{8}$\ $cm^{-1}$ and $C_{pt}=3.8\times 10^{-13}$ $MeVb$.
Furthermore, $g_{e}=10$ and $r=1.7\times 10^{-11}\exp \left[ 2\pi \eta
\left( E\right) \right] $ where $\eta \left( E\right) =\alpha _{f}\sqrt{%
m_{0}c^{2}/E}$ resulting $r=1$ at $E=3.186$ $keV$. From this number one can
conclude that the yield produced by the electron assisted $d+d\rightarrow p+t
$ process with decreasing beam energy\ becomes comparable with and larger
than the yield produced by the normal $d+d\rightarrow p+t$ process near and
below $3$ $keV$. Since $C_{nHe}=3.0\times 10^{-13}$ $MeVb$ therefore $\sigma
_{nHe}$ has the same order of magnitude as $\sigma _{pt}$ has and so similar
statement can be made in the case of electron assisted $d+d\rightarrow n+$ $%
^{3}He$ reaction too. Consequently, one can conclude that the electron
assisted $d+d\rightarrow p+t$ and $d+d\rightarrow n+$ $^{3}He$ processes
should be taken into account when evaluating the data of low energy fusion
reactions in metals.

\end{document}